\documentclass[prl,twocolumn,showpacs,floatfix,preprintnumbers,amsmath,amssymb,superscriptaddress]{revtex4}

\usepackage{graphicx}
\usepackage{amsmath}
\usepackage{amssymb}
\usepackage{color}
\usepackage{dcolumn}
\usepackage{epsfig}
\usepackage{bm}
\usepackage[urlcolor=blue]{hyperref}
\hypersetup{backref, colorlinks=true, linkcolor=blue, citecolor=blue}

\newcommand{\nef}{N(0)}

\begin{document}

\preprint{}

\title{{High pressure effects on isotropic superconductivity in the iron-free layered pnictide superconductor BaPd$_2$As$_2$}}

\author{Mahmoud Abdel-Hafiez}
\email{mahmoudhafiez@gmail.com}
\affiliation{Center for High Pressure Science and Technology Advanced Research, Beijing, 100094, China}
\affiliation{Institute of Physics, Goethe University Frankfurt, 60438 Frankfurt/M, Germany}
\affiliation{National University of Science and Technology "MISiS", Moscow 119049, Russia}

\author{Y. Zhao}
\affiliation{Center for High Pressure Science and Technology Advanced Research, Shanghai, 201203, China}

\author{Z. Huang}
\affiliation{Department of Physics, The Hong Kong University of Science and Technology, Clear Water Bay, Kowloon, Hong Kong}

\author{C.-w. Cho}
\affiliation{Department of Physics, The Hong Kong University of Science and Technology, Clear Water Bay, Kowloon, Hong Kong}

\author{C. H. Wong}
\affiliation{Institute of Physics and Technology, Ural Federal University, Yekaterinburg, Russia}

\author{A. Hassen}
\affiliation{Faculty of science, Physics department, Fayoum University, 63514-Fayoum- Egypt}

\author{M. Ohkuma}
\affiliation{Graduate School of Engineering, Kyushu Institute of Technology, Kitakyushu 804-8550, Japan}

\author{Y.-W. Fang}
\affiliation{Department of Electronic Engineering, East China Normal University, Shanghai 200241, China}

\author{B.-J. Pan}
\affiliation{Institute of Physics and Beijing National Laboratory for Condensed Matter Physics, Chinese Academy of Sciences, Beijing 100190, China}

\author{Z.-A. Ren}
\affiliation{Institute of Physics and Beijing National Laboratory for Condensed Matter Physics, Chinese Academy of Sciences, Beijing 100190, China}

\author{A. Sadakov}
\affiliation{P.N. Lebedev Physical Institute, Russian Academy of Sciences, 119991 Moscow, Russia}

\author{A. Usoltsev}
\affiliation{P.N. Lebedev Physical Institute, Russian Academy of Sciences, 119991 Moscow, Russia}

\author{V. Pudalov}
\affiliation{P.N. Lebedev Physical Institute, Russian Academy of Sciences, 119991 Moscow, Russia}


\author{M. Mito}
\affiliation{Graduate School of Engineering, Kyushu Institute of Technology, Kitakyushu 804-8550, Japan}

\author{R. Lortz}
\affiliation{Department of Physics, The Hong Kong University of Science and Technology, Clear Water Bay, Kowloon, Hong Kong}

\author{C. Krellner}
\affiliation{Institute of Physics, Goethe University Frankfurt, 60438 Frankfurt/M, Germany}

\author{W. Yang}
\affiliation{Center for High Pressure Science and Technology Advanced Research, Shanghai, 201203, China}
\affiliation{High Pressure Synergetic Consortium (HPSynC), Carnegie Institution of Washington, Argonne, Illinois, 60439, USA.}

\date{\today}

\begin{abstract}
While the layered 122 iron arsenide superconductors are highly anisotropic, unconventional, and exhibit several forms of electronic orders that coexist or compete with superconductivity in different regions of their phase diagrams, we find in the absence of iron in the structure that the superconducting characteristics of the end member BaPd$_2$As$_2$ are surprisingly conventional. Here we report on a complementary measurements of specific heat, magnetic susceptibility, resistivity measurements, Andreev spectroscopy and synchrotron high pressure X-ray diffraction measurements supplemented with theoretical calculations for BaPd$_2$As$_2$. Its superconducting properties are completely isotropic as demonstrated by the critical fields, which do not depend on the direction of the applied field. Under the application of high pressure, $T_{c}$ is linearly suppressed, which is the typical behaviour of classical phonon-mediated superconductors with some additional effect of a pressure-induced decrease in the electronic density of states and the electron-phonon coupling parameters. Structural changes in the layered BaPd$_2$As$_2$ have been studied by means of angle-dispersive diffraction in a diamond-anvil cell.  At 12\,GPa and 27.2\,GPa we observed pressure induced lattice distortions manifesting as the discontinuity and, hence discontinuity in the Birch-Murnaghan eqation of state. The bulk modulus is $B_{0}$ = 40(6)\,GPa below 12\,GPa and $B_{0}$ = 142(3)\,GPa below 27.2\,GPa.
\end{abstract}


\maketitle



\section{I. Introduction}

In conventional superconductors, the electron-phonon interaction gives rise to the attraction between electrons near the Fermi-surface with opposite momenta and opposite spins, which eventually causes superconductivity and conservation of the time-reversal symmetry~\cite{SC}. The discovery of superconductivity up to 55\,K in iron-based pnictides has been at the forefront of interest in the scientific community over the last few years~\cite{Kam,XHC,pag,Hirschfeld}. These materials have multiple Fermi pockets with electronlike and holelike dispersion of carriers. For BaFe$_{2}$As$_{2}$ (BFA) parent compound, the electron doping is induced by substitution of Ni, Co, Rh, and Pd at the Fe sites, which have more $d$-electrons than Fe~\cite{1,2,5}. The parent compound BFA itself is not superconducting, but features two forms of highly correlated electronic orders. One is the antiferromagnetic spin density wave (SDW) of stripe-type order in which chains of parallel spins are adjacent to chains with opposite spin direction~\cite{Chu}. In addition to this SDW order a spontaneous breaking of rotational symmetry (nematic order) from a high-temperature tetragonal (C$_{4}$) symmetry to a low temperature orthorhombic structure occurs~\cite{Chu} . The nematic order is most likely due to an electronic instability that causes pronounced in-pane anisotropy of the Fe-As layers~\cite{Chu2}. Upon doping, both forms of static orders coexist with superconductivity in a wide range of their phase diagram and it is still unclear whether they are beneficial or competing with superconductivity. The compound BaPd$_2$As$_2$ in the focus of this article represents the end member of the Pd-doped series and exists in the form of two types of crystal structures: ThCr$_{2}$Si$_{2}$-structure type (I4/mmm) and CeMgSi$_{2}$ -structure type (P4/mmm)~\cite{23,24}. The former structure has bulk superconductivity with $T_{c}$ = 3.85\,K, while in the latter structure only filamentary superconductivity was observed below 2\,K. This shows that the crystal structure has a predominant effect on the superconducting properties of these systems~\cite{25,26,27}. Given that Fe likely has a leading effect on the high temperature superconductivity in doped BFA, the effect of partial substitution of Fe on superconducting properties in general, and on their anisotropy evolution remains unaddressed. The heavily hole-doped compounds  differ from their optimally doped counterparts by the presence of particularly strong electronic correlations~\cite{K1}, which is seen in their substantially larger reported mass enhancements compared to those of the optimally doped compounds. For instance, it is well established that CaFe$_{2}$As$_{2}$ is less correlated than KFe$_{2}$As$_{2}$ at ambient pressure~\cite{K2,K3}. Therefore, it is necessary to deeper investigate the evolution of physical properties of the BFA system under various doping, including  end members of the 122 family with fully substituted Ba, Fe, and As.

In particular, due to the lack of systematic experiments, a comprehensive and complete picture on the iron-free layered pnictide superconductor BaPd$_{2}$As$_{2}$ is still lacking. Here, we performed thermodynamic, transport, magnetotransport, local spectroscopy measurements, and first principle calculations in order to unravel the nature of the superconductivity in the BaPd$_{2}$As$_{2}$ system. Specifically, we examined the electronic properties of BaPd$_2$As$_2$ single crystal using a combination of specific heat (down to 400\,mK), magnetic susceptibility, resistivity measurements (down to 400\,mK and under pressure up
to 11GPa), Andreev reflection spectroscopy (down to 1.4\,K), and first-principles calculations. We found that BaPd$_2$As$_2$ as the end member of the substituted series of BaFe$_{2-x}$Pd$_{x}$As$_{2}$ appears to behave surprisingly different from the parent compound BaFe$_2$As$_2$. While highly unconventional superconductivity can be induced by doping or application of pressure in BaFe$_2$As$_2$, the end member BaPd$_2$As$_2$ in which all iron is completely replaced by Pd, appears to be a very conventional classical $s$-wave superconductor most probable, with  phonon-mediated pairing. All Fe-based 122 superconductor materials are layered compounds, and are expected to show highly anisotropic superconducting parameters. While such anisotropy is indeed found in the iron based superconductors, the superconducting properties of BaPd$_2$As$_2$ are isotropic. This demonstrates the important role of the iron in the iron-based superconductors, which causes all the unconventional effects, including the competing or coexisting orders such as spin density waves and nematic orders, that appear to be completely absent in BaPd$_2$As$_2$.

\section{II. Experimental}

Single crystals of ThCr$_{2}$Si$_{2}$-type BaPd$_2$As$_2$ were successfully prepared by a self-flux method, details for the  growth process and sample characterization were published elsewhere\cite{24}. Low temperature transport and specific heat (down to 400 mK) were measured with a Physical Property Measurement System (PPMS, Quantum Design). Four contacts were used to measure the high-pressure in-plane resistivity with the superconductor BaPd$_{2}$As$_{2}$ set in a diamond anvils cell in the PPMS-9\,T. Single crystal was loaded in the sample space formed by c-BN gasket of around 130\,$\mu m$ in diameter, made of a miniature diamond anvil cell. We used insulating gasket made of the mixture of cubic boron nitride with epoxy. Daphne oil 7373 was used as a pressure medium. Pressure was calibrated by using several ruby chips with dimensions of about 1\,mm were placed into the cell along with the sample at room temperature\cite{Ru}. Four Pt leads with thickness of about 1\,$\mu m$ and the width of about 7-12\,$\mu m$ were used for 4-probe measurements. The DC magnetization was measured in the Quantum Design SQUID VSM magnetometer with magnetic fields applied parallel and perpendicular to the $ab$ plane. For the parallel field configuration the sample was attached with vacuum grease to a quartz sample holder, which allowed us to rotate the sample at room temperature in the grease so that different in-plane field orientations could be realized. AC magnetic susceptibility measurements were performed with an AC field at a frequency of 10 Hz and amplitude of 3.86 Oe over the pressure range of 0 - 5\,GPa in a commercial SQUID magnetometer.  Pressure was attained by a miniature diamond anvil cell (DAC) which was desigend to be inserted into the SQUID magnetometer~\cite{27_1}. In the sample chamber, the crystals were immersed into a pressure transmitting medium, Apiezon J oil, together with a piece of ruby as the manometer. Pressure calibration was performed using the ruby fluorescence method at room temperature\cite{Ru}. Multiple Andreev reflection effect (MARE) spectroscopy of superconductor - normal metal - superconductor (SnS) junctions was performed using the break-junction technique~\cite{V1,V2}. MARE occurs in the ballistic regime of SnS junctions and causes excess current at low bias voltages (so-called foot) and a subharmonic gap structure. This structure consists of series of dips in dynamic conductance, each dip corresponding to voltage values $V_{n}$=2$\Delta$/$en$, where $\Delta$ is the superconducting gap, $e$ is the elementary charge, and $n$ = 1, 2, . . . is the subharmonic order. First-principles calculations are performed with plane-wave density functional theory (DFT) implemented in Quantum Espresso~\cite{27_2}. The in situ high-pressure X-ray diffraction ( $\lambda$ = 0.4066 \AA) measurement was performed with an angle-dispersive synchrotron X-ray diffraction mode (AD-XRD) at beamline 16 IDB of the Advanced Photon Source, Argonne National Laboratory. The as-prepared single crystal sample was crashed to powder and loaded into a gasketed diamond anvil cell (DAC) with silicon oil as a pressure-transmitting medium up to 30.3 GPa.

\section{III. Results and discussion}

\begin{figure}
\includegraphics[width=20pc,clip]{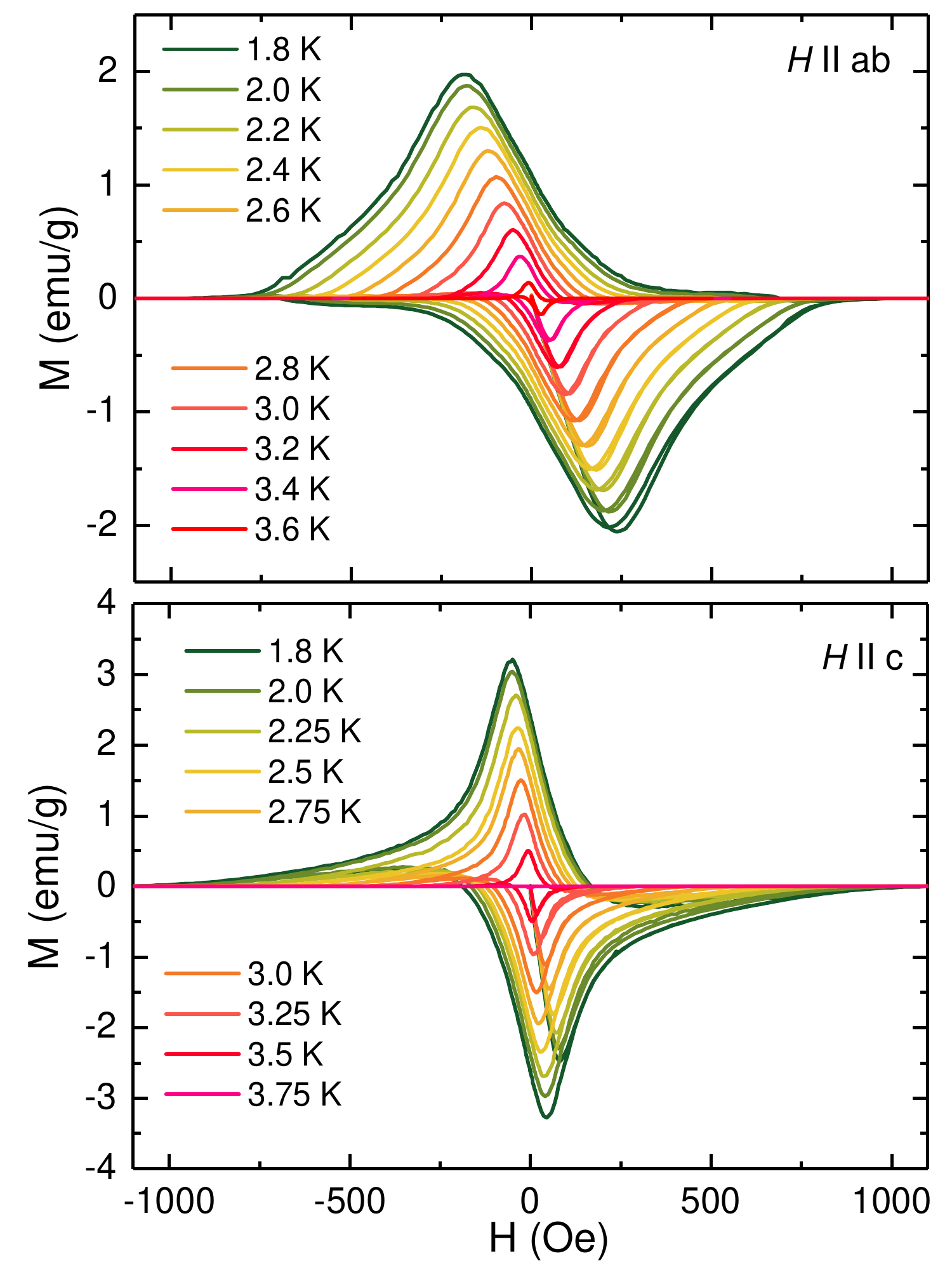}
\caption{(a) and (b) summarize the isothermal magnetization $M$ Vs. $H$ loop up to 3.6\,K for $H \parallel ab$ and up to 3.75\,K for $H \parallel c$, respectively.}
\end{figure}

\begin{figure}
\includegraphics[width=20pc,clip]{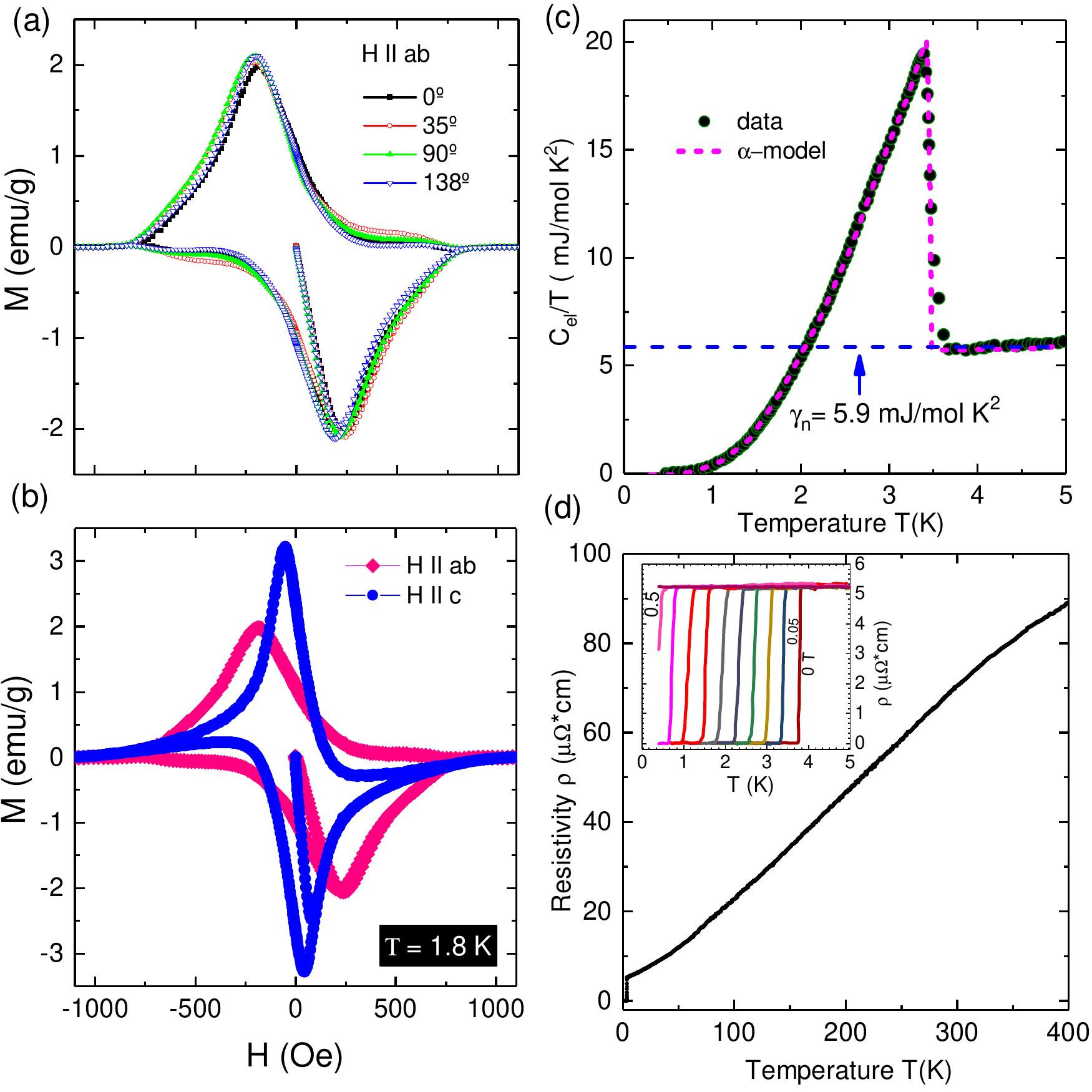}
\caption{(a) represents the identical values of magnetic moment with different in-plane angles. (b) shows the isothermal magnetization $M$ vs. $H$ loop at 1.8\,K for $H \parallel ab$ and  $H \parallel c$. (c) illustrates the superconducting electronic specific heat $c_{el}$ of BaPd$_2$As$_2$ as a function of the temperature (after subtracting the phonon contribution). The dashed line represents the theoretical curve based on single-band weak coupling BCS theory based on the $\alpha$-model. (d) presents the in-plane electrical resistivity $\rho$ of BaPd$_2$As$_2$ versus temperature. the inset shows the superconducting transition for different values of applied field (0T, 0.05T up to 0.5T) down to 400\,mK.}
\end{figure}

\begin{figure}
\includegraphics[width=20pc,clip]{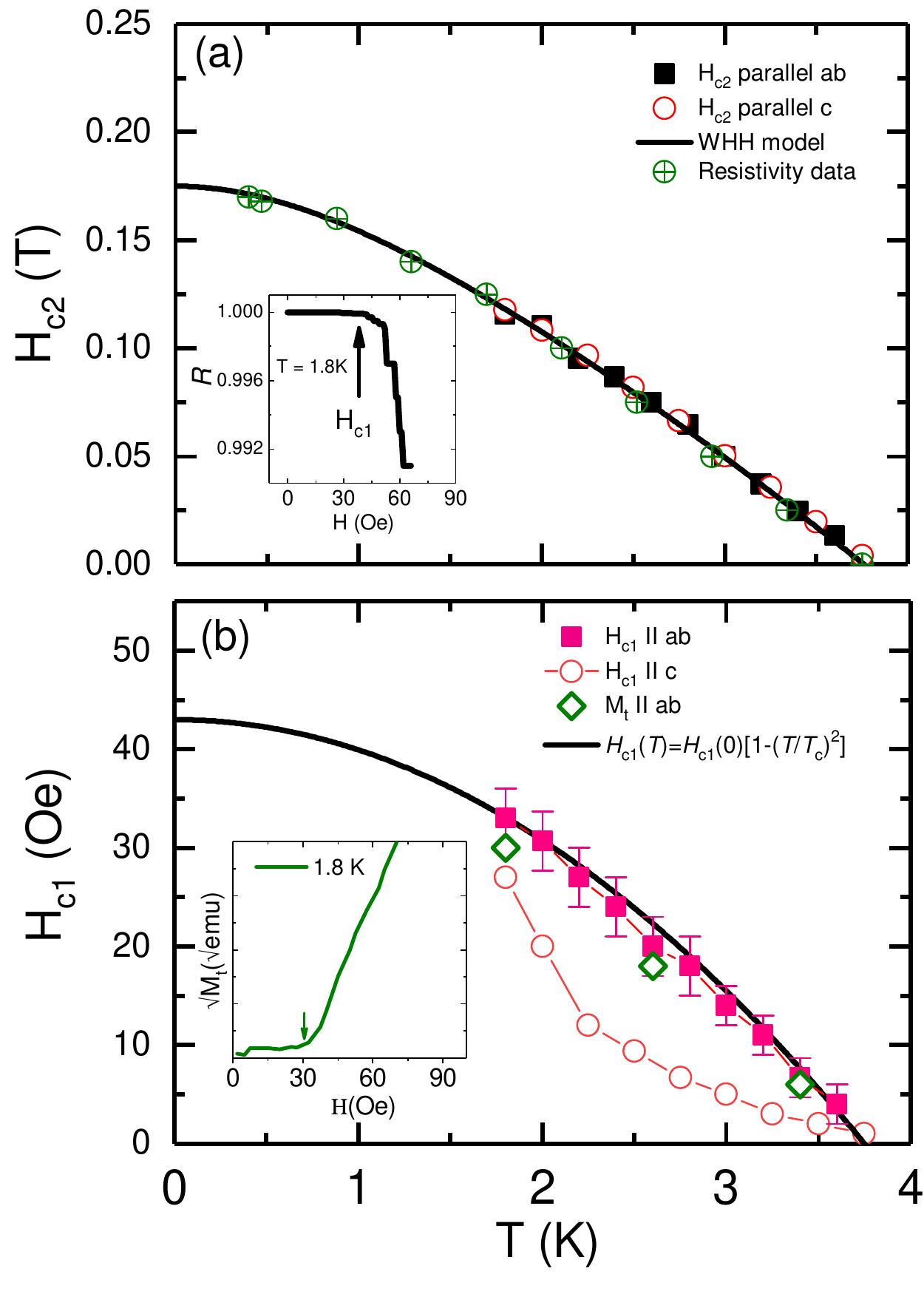}
\caption{(a) The upper critical field $H_{c2}$ vs. temperature for $H \parallel ab$ and $H \parallel ab$ with the best fit to the experimental data by the WHH model. (b) Phase diagram of $H_{\mathrm{c1}}$ vs. $T$ for the field applied parallel to $c$ axis. $H_\mathrm{c1}$ has been estimated by two different methods - from the extrapolation of $\sqrt{M_{t}}\rightarrow 0$ (open symbols, see lower inset) and from the regression factor (closed symbols, see the inset of (a)). The bars show the uncertainty of estimated by the deviating point of the regression fits and the linear fit of $\sqrt{M_{t}}$.}
\end{figure}

\subsection{A. Thermodynamic studies}

The isothermal magnetization $M$ vs. $H$ loops $H \| ab$ and $H \| c$ at different temperatures up to 3.75\,K is shown in Fig.\,1a and 1b, respectively. The fact that the hysteresis loops for both orientations are point symmetric at the origin points to relatively weak surface barriers, and thus is indicative of bulk pinning (Fig.\,1a-b). This consideration applies to all data up to $T_{c}$ and ensures that vortex penetration occurs at a field near the lower critical field, $H_{c1}$. In contrast, when surface barriers were predominant, the first vortex entrance would take place at much higher field ($\approx H_{c}$). This is a very important point to obtain reliable estimation of the thermodynamic lower critical field, as discussed below. It is worth noting that the superconducting hysteresis loops can still be measured at temperatures very close to $T_{c}$ with only a weak magnetic background. This indicates that the sample contains negligible magnetic impurities. The bulk superconducting transition temperature $T_{c} \approx$ 3.5\,K as determined by specific-heat measurements (Fig.\,2c) is significantly higher than $T_{c}$ = 1.27 and 0.92\,K for CaPd$_2$As$_2$ and SrPd$_2$As$_2$, respectively~\cite{23}. The jump height of the specific heat at $T_c$ is found as  $\Delta C_{el}/T_{c}$ $\approx$ 13.0\,mJ/mol K$^{2}$. From our obtained $\gamma_n$ (known as the Sommerfield coefficient in the normal state) = 5.9\,mJ/mol K$^{2}$ value, we estimate the universal parameter $\Delta C_{el}/\gamma_nT_c$ = 2.2, which is somewhat greater than the weak-coupling BCS prediction of 1.43 and indicates superconductivity in the strong-coupling regime. The excellent fit with an isotropic single-band $s$-wave alpha model~\cite{27_3} (Fig.\,2c) suggests a conventional nature a single-band of superconductivity in BaPd$_2$As$_2$. From the relation
of the Debye temperature, \emph{$\theta_{D}$ = (12$\pi$ $^{4}$RN/5$\beta$) $^{1/3}$}, where \emph{$R$} is the molar gas constant and \emph{$N$} = 5 is the number of atoms per formula unit, we obtain
\emph{$\theta_{D}$} = 144\,K. Table 1 shows a comparison of the specific heat parameters of several \emph{A}Fe$_{2}$As$_{2}$ compounds where \emph{A} is a divalent atom and BaPd$_2$As$_2$ system. Clearly, a strongly enhanced value $\gamma_n$ is present for the heavily hole doped compound KFe$_2$As$_2$ compared to other stoichiometric 122 compounds. However, several theoretical considerations, including two proposed modified Kadowaki-Woods relations~\cite{KW1,KW2}, as well as the  observation of a significant linear in temperature residual term, point to a significantly smaller value of about 60 mJ/mol K$^{2}$ for the Sommerfeld coefficient for the itinerant quasiparticles. This suggests that the strongly correlated "heavy-fermion-like" scenario suggested for KFe$_{2}$As$_{2}$ in the literature should be revisited\cite{U2,M3}. On the other hand, the low values of $\gamma _{n}$ for BaFe$_{2}$As$_{2}$ and SrFe$_{2}$As$_{2}$ are not surprising since large parts of its Fermi surface are gapped due to the well-known magnetic spin density wave (SDW) transition at high temperatures.
\begin{center}
\begin{table}[h]
\caption{Comparison of the specific heat parameters of several
stoichiometric AFe$_{2}$As$_{2}$ compounds, where A is a divalent
atom and BaPd$_2$As$_2$ system. $\gamma_n$ is given in units (mJ/mol K$^{2}$) and $\beta$ is
given in units (mJ/mol K$^{4}$).}
\centering
\begin{tabular}{@{}*{7}{l}}
\textbf{ Compound} & \textbf{$\gamma$} & \textbf{$\beta$} & \textbf{$\theta$$_{D}$(K)} & \textbf{Ref.}\\ \hline
KFe$_{2}$As$_{2}$ & 74 (2) & 0.71  & 239 & \cite{U2,M3} \\
CaFe$_{2}$As$_{2}$ & 8.2(3) & 0.383  & 292 & \cite{Ronning2008,R3}\\
SrFe$_{2}$As$_{2}$& 33  & 0.64 & 248 & \cite{Yan2008} \\
BaFe$_{2}$As$_{2}$& 6.1, 37 & 1.51, 0.6 & 186, 250 & \cite{Dong2008,Kreyssig2008,M5}\\
BaPd$_{2}$As$_{2}$& 5.9 & 1.16 & 144 & This study,\cite{Mc7} \\\hline
\end{tabular}
\end{table}
\end{center}

Figure 2d illustrates the temperature dependence of the in-plane electrical resistivity at ambient pressure. It exhibits a metallic behavior in the normal state followed by a sharp superconducting transition at $T_c=3.8 - 3.85$\,K with $\Delta T_{c}$ = 0.15\,K. The residual resistivity ratio (RRR) is found to be 18. The observations of a large RRR and a narrow SC transition again indicate a high quality of the samples investigated here. The inset of Fig.\,2d shows an enlarged view at the superconducting transition which summarizes the $T$-dependent resistivity measured in various magnetic fields. Under a magnetic field  the superconducting transition is shifted significantly to a lower temperature.

\subsection{B. H - T phase diagram}

The Cooper pairs are destroyed by the application of a high magnetic field either by orbital pair breaking due to the Lorentz force or by Pauli paramagnetic pair breaking via the Zeeman effect. Critical fields are one of the fundamental parameters that provide valuable information about the microscopic origin of pair breaking, and reflect the electronic structure responsible for superconductivity. Figure.\,2a shows the isothermal magnetization of BaPd$_2$As$_2$ measured at various in-plane magnetic field orientations ($H \parallel ab$) at different in-plane angles with respect to the crystalline axes. The data provides clear evidence that all data converges to the same upper critical field value, indicating that this system is isotropic in the $ab$ plane. The isothermal magnetization for both $H \parallel ab$ and $H \parallel c$ axis is presented in Fig.\,2b, in which the upper critical field $H_{c2}$ (field at which $M$= 0) also converge to the same value. Note that the differences in the magnetization loops here are attributed to the demagnetization factor of the rather flat sample geometry.

In order to illustrate the upper critical field $H_\textup{c2}$, we show the magnetic phase diagram in Fig.\,3(a). The Werthamer-Helfand-Hohenberg (WHH) theory predicts the behavior of $H_\mathrm{c2}(T_\mathrm{c})$, taking into account paramagnetic and orbital pair-breaking~\cite{NR}. The orbital limited field $H_{\mathrm{c2}}^{\mathrm{orb}}$ at zero temperature is determined by the slope at $T_\mathrm{c}$ as $H_{\mathrm{c2}}^{\mathrm{orb}} = 0.69 \,T_{\mathrm{c}}\, (\partial H_{\mathrm{c2}}/\partial T) |_{T_{\mathrm{c}}}$. Fit to the data in the entire measurement range for negligible spin-paramagnetic effects ($\alpha = 0$) and spin-orbit scattering ($\lambda = 0$) yields $\mu_0 H_{\mathrm{c2}}^{\mathrm{orb}}(B \parallel c)$ = 0.18~T for both orientations. The most remarkable aspect of Fig.\,3a is the fact that the upper critical field extrapolates to a similar zero-temperature value (0.18\,T), irrespective of whether the field is applied parallel or perpendicular to the $c$-axis. This is in contrast to the behaviour of other quasi-two dimensional superconductors~\cite{U1,U2,U3,U4}, where the in-plane critical fields are many times larger than those for fields applied perpendicular to the quasi-two dimensional planes.

To further study the anisotropy of the superconducting state, we have measured temperature dependence of the lower critical field, $H_{c1}$, (see Fig.\,3b). The lower critical field $H_{c1} (T)$, where the penetration of vortices in the sample becomes energetically favorable, and the magnetic penetration depth $\lambda (T)$ are very useful parameters that provide important information about bulk thermodynamic properties. However, the determination of $H_{c1}$ from magnetization data in not an easy task. These types of data are obtained by tracking the virgin $M(H)$ curve in low fields at several temperatures. We have adopted a rigorous procedure (i.e., with a user-independent outcome) to determine the transition from linear to nonlinear $M(H)$, that consists of calculating the regression coefficient $R$, see the inset of Fig.3a, of a linear fit to the data points collected between 0 and $H$, as a function of $H$~\cite{m1,m2,m3}. In order to trust our extracted $H_{c1}$ values, we determined the $H_{c1}$ for some particular temperatures by measuring the onset of the trapped moment $M_{t}$, see the lower inset of Fig.\,3b. The trapped flux moment $M_{t}$ is obtained by (i) warming the sample up to temperatures above $T_{c}$, then (ii) cooling the sample at zero field down to each particular temperature, subsequently (iii) the external magnetic field is increased to a ceratin maximum value $H_{m}$ and (iv) measure the remanent magnetization $M_{t}$ after the applied field has been switched off. The field $H_{m}$ at which $M_{t}$ deviates from zero determines the $H_{c1}$ value at the desired temperature. The so obtained values of $H_{c1}$ are shown in Fig.\,3b for  $H \parallel ab$.

Although the low-temperature lower critical field is rather isotropic, the initial slope near $T_{c}$ does show some dependence on the field orientation (Fig.\,3b), perhaps resulting from details of the vortex structure, the Fermi-surface topology or different sample edge properties. $H_{c1}$ clearly flattens off at low temperatures indicating that the superconducting energy gap becomes fully open. In the limit of $T$ $\approx$ 0\,K, we acquired the value of $H_{c1}$ = 43 $\pm$ 2\,Oe. From the measured upper critical magnetic field, we can estimate the effective coherence length $\xi$ =  $\sqrt{\frac{\phi_{0}}{2\pi H_{c2}}}$ where $\phi_{0}$ is the flux quantum, giving $\xi$ = 42.7 $\pm$ 0.5\,nm. Exploiting the following expressions relating $H_{c1}$ and $H_{c2}$ derived by Brandt~\cite{BR}: $\frac{2H_{c1}(0)}{H_{c2}(0)}$ = $\frac{\ln\kappa + \delta(\kappa)}{2\kappa^{2}}$ =  $\frac{\ln\kappa + 0.5}{\kappa^{2}}$ provides us with the effective Ginzburg -Landau parameter $\kappa$ = 0.62 $\pm$ 0.01 along with the effective penetration depth $\lambda$ = $\kappa \xi$ = 26.4 $\pm$ 1\,nm.

\subsection{C. Andreev reflections effect spectroscopy}

\begin{figure}
\includegraphics[width=21pc,clip]{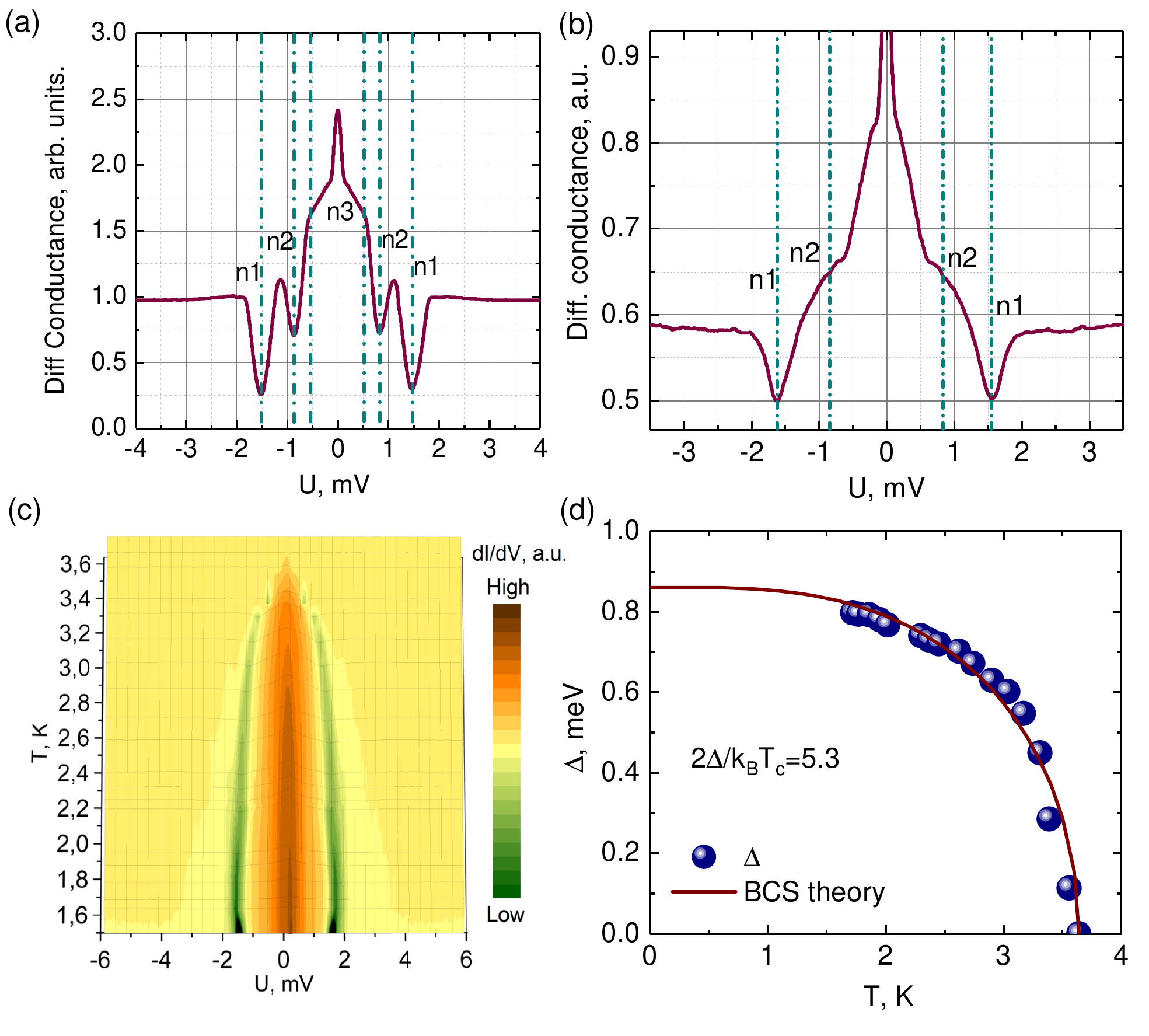}
\caption{(a) Dynamic conductance spectrum measured at $T$ = 1.63\,K for BaPd$_{2}$As$_{2}$ single crystal with local critical temperature $T_{c}$=3.75\,K and for SnS Andreev contact with 3 Andreev reflections. (b) Dynamic conductance spectrum with 2 Andreev reflections at $T$ = 1.63 K. Vertical green lines on panels a and b depict the  dip positions. (c) Temperature evolution of conductance from the latter junction. (d) Temperature dependence of the obtained superconducting gap. Solid line shows a BCS-like curve for $\delta$.}
\end{figure}

\begin{figure}
\includegraphics[width=21pc,clip]{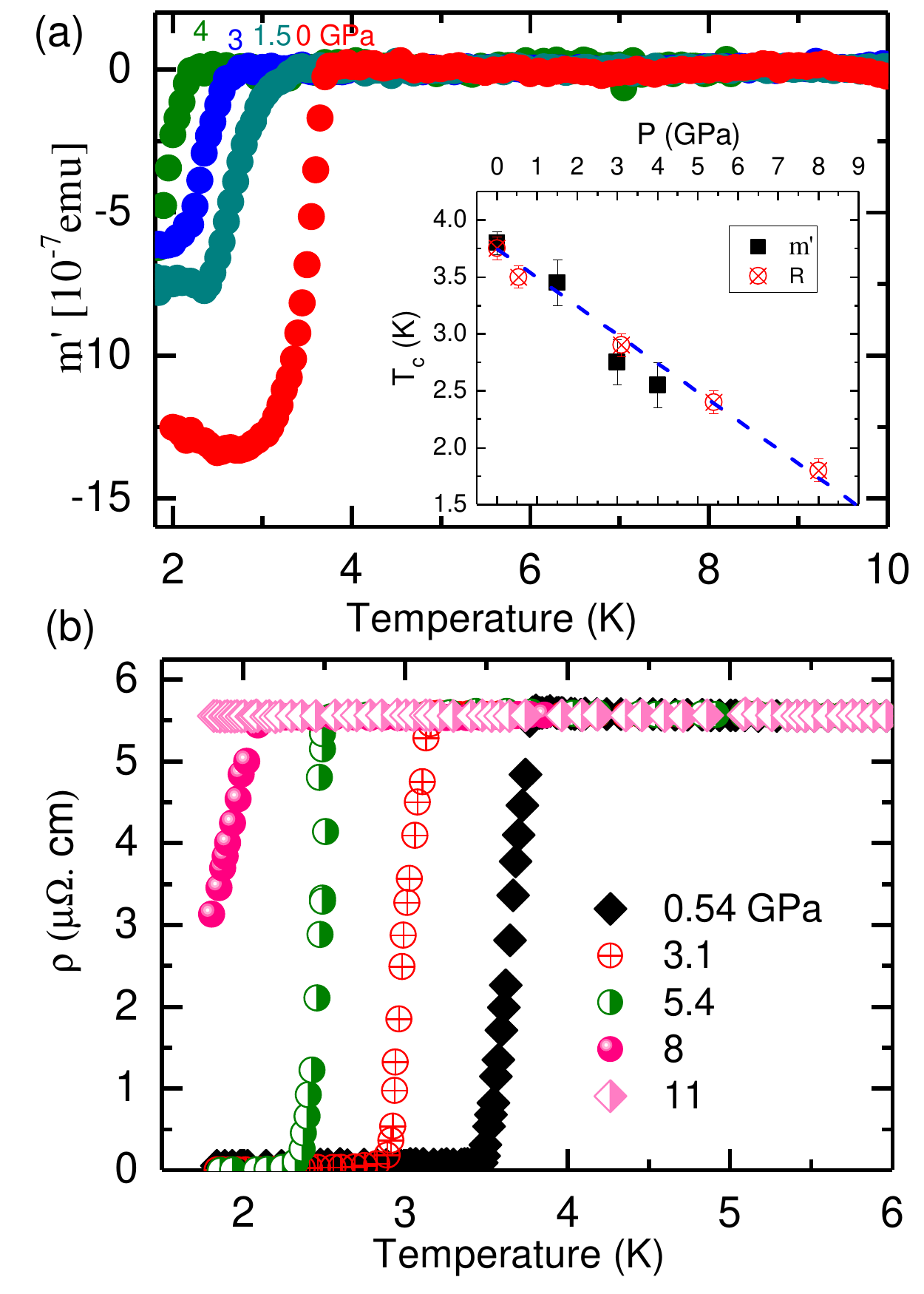}
\caption{Temperature dependence of magnetic moment of at different values of applied pressure. The inset shows the P-T pase diagram which illustrates the decrease of $T_{c}$ with the applied pressure. (b) In-plane electrical resistivity of BaPd$_2$As$_2$ versus temperature showing the SC transition for different values of the applied pressure.}
\end{figure}

\begin{figure}
\includegraphics[width=21pc,clip]{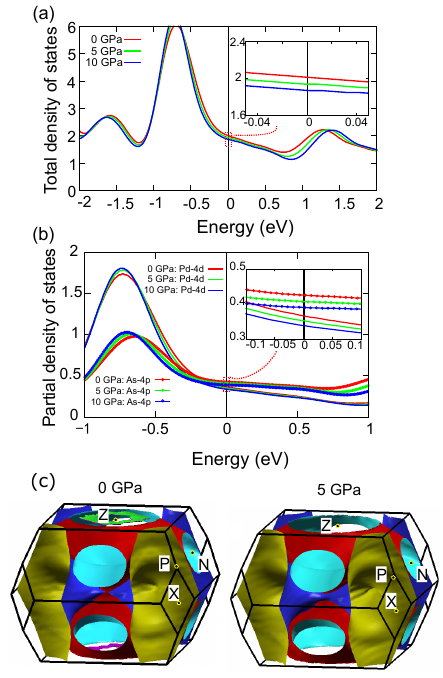}
\caption{(a) Total density of states (states/eV) of BaPd$_2$As$_2$~under pressures. (b) Pd-4$d$ and As-4$p$~partial densities of states (states/eV) under pressures.
  (c) The 3D Fermi surfaces in the Brillouin zone under pressures of 0 and 5 GPa. BaPd$_2$As$_2$ shows multi-band Fermi surface.}
\end{figure}

In order to further test the superconductivity character in the investigated system, we performed multiple Andreev reflections effect spectroscopy. The latter is known to be a powerful tool for probing the superconducting energy gaps. Several adjustable SnS junctions were made  using a break-junction (BJ) technique; for details of the BJ techniques, see, e.g.~\cite{A1,A2,A3}. Firstly, the $dI/dV$ curves for the junctions clearly demonstrate an enhanced conduction at zero bias, intrinsic to Andreev-type SnS junctions (see Figs.\,4a, and b). Secondly, the $dI/dV$ curves in Figs.\,4a, and b show a series of sharp dips located at  bias voltages $V \propto 1/n$ with $n=1, 2, 3$. Several prepared contacts demonstrated similar spectra with up to three sharp dips (see Figs 3a and b), whose positions were reproducible in various spectra within $\approx 0.05$meV.

For these reasons, we interpret the series of dips as the subharmonic gap structure, typical for the multiple Andreev reflection effect; correspondingly, we associate the gap positions $V_n$ with $2\Delta/$en~\cite{A1,A2}. Having established this,  we find $\Delta=0.80$meV from the spectra Fig.\,4a, and $\Delta=0.81$meV from Fig.\,4b. Temperature dependence of the Fig.\,4a spectrum is shown in Fig.\,3c. As temperature increases, the voltage distance between the two strips in Fig.\,4c shrinks, signifying gradual decay of the gap. The resulting  temperature dependence of the energy gap, $\Delta(T)$, is plotted in Fig.\,4d; it almost does not deviate  from the single-band BCS like curve. The gap vanishes to zero at a local $T_c=3.64$\,K, very close to that found from $H_{c1}(T)$ and $H_{c2} (T)$ measurements. By extrapolating $\Delta(T)$ to zero temperature we obtain an estimate $\Delta(0)\approx$ 0.85\,meV. With it, the characteristic BCS ratio $2\Delta/k_BT_c = 5.4$, which is rather close to the estimate  $2\Delta/k_BT_c = 4.76$, obtained from the specific heat measurements (Fig.\,2c), again indicating the strong coupling case.

\begin{figure}
\includegraphics[width=21pc,clip]{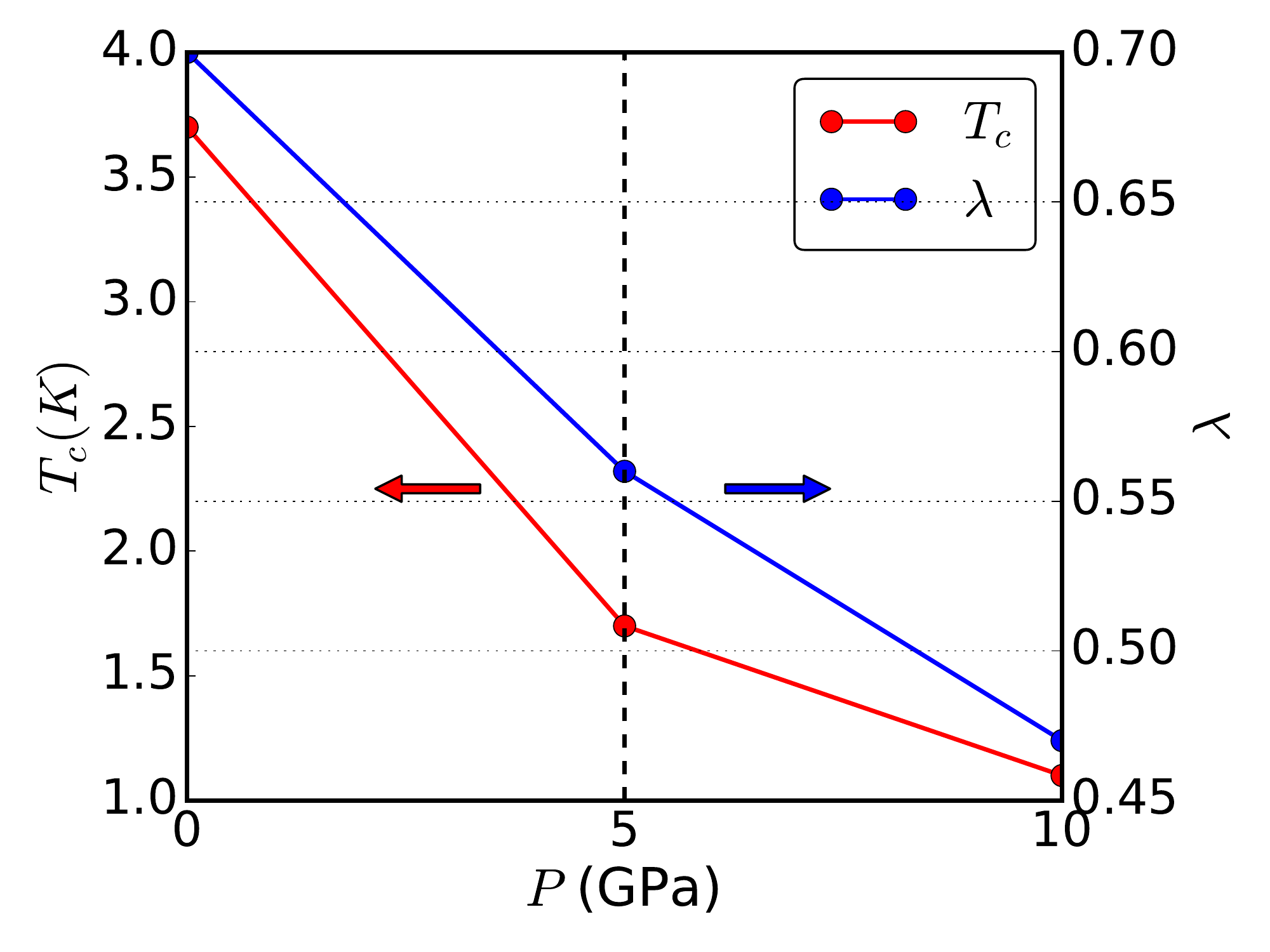}
\caption{ The pressure dependence of electron phonon coupling parameter $\lambda$ (right vertical axis) and the  superconducting transition temperature $T_{c}$ (left vertical axis) from first-principles study.}
\end{figure}

\subsection{D. High-pressure studies}

\begin{figure*}[tbp]
\includegraphics[width=37pc,clip]{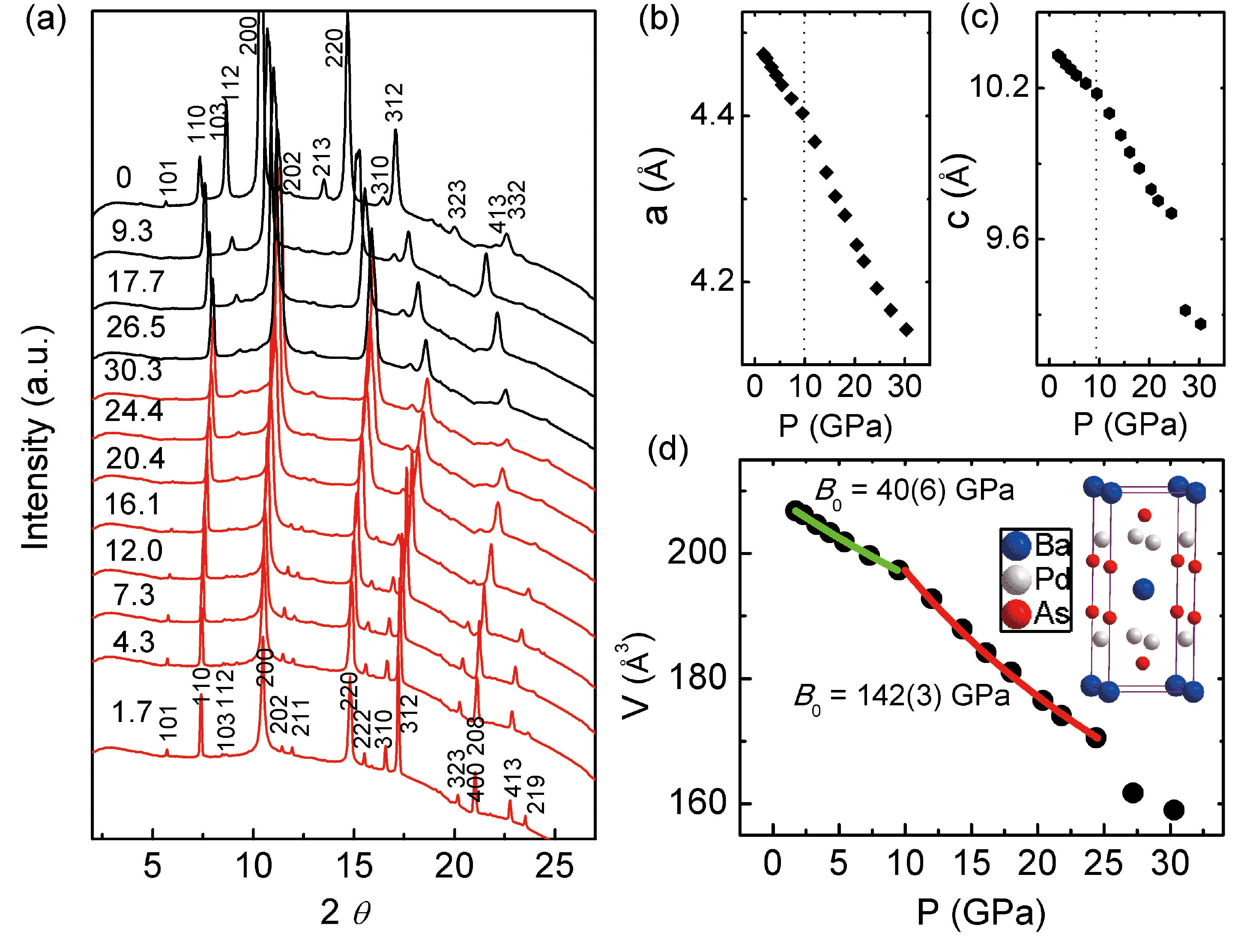}
\caption{\label{fig:wide}  (a) Angle-dispersive XRD patterns of BaPd$_{2}$As$_{2}$ at selected pressures. The lattice parameter are shown in (b) and (c), obviously, both of the parameters a and c decrease faster with pressure above 9.5 GPa, while c shows a discontinue at 27.2\,GPa. In (a) The upper black patterns represents data at increasing pressure, while the lower red ones shows data taken at decompression. (d) Unit cell volume obtained from Rietveld refinement of powder XRD patterns as a function of pressure. The inset shows its crystal structure.}
\end{figure*}

Another important result of our experiment is the effect of pressure on BaPd$_2$As$_2$ system. Pressure has long been recognized as a fundamental thermodynamic variable, which is a convenient tool for deep understanding of various SC characteristics. It is considered as a clean way to tune basic electronic and structural properties without changing the stoichiometry. In 122 iron-based superconductors, it was found that $T_{c}$ and upper critical field of the parent undoped compound KFe$_{2}$As$_{2}$~\cite{31} have a V shaped dependence on pressure~\cite{32}. In BaFe$_2$As$_2$, a pressure-induced structural distortion occurs~\cite{33}. To investigate BaPd$_2$As$_2$ under pressure, we performed AC susceptibility and electric transport experiments in a diamond anvil cell. In Fig.\,5, we show data of the AC susceptibility (Fig.\,5a) and the electrical resistivity (Fig.\,5b), both of which show a gradual suppression of $T_{c}$ with pressure, with excellent agreement between the two methods. The in-plane electrical resistivity at different pressures up to 11\,GPa (Fig.\,5b) shows that superconductivity is completely suppressed at 11\,GPa. On the basis of the above results, we plotted a temperature-pressure phase diagram for BaPd$_2$As$_2$ in the inset of Fig.\,5a. The linear decrease of $T_{c}$ under pressure is the typical behavior found in classical superconductors, where phonon hardening is the most probable effect which usually reduces the critical temperature and ultimately completely suppresses it. In the following, we investigate whether this classical scenario holds here, or whether there are other factors that regulate the $T_{c}$ suppression.

In Fig.\,6, we present the calculated electronic structures of BaPd$_2$As$_2$ under various pressures (0, 5, and 10 GPa). The crystal structures under these pressures are fully optimized within the framework of DFT. In order to show that our calculations can reproduce the crystal structures, we compare our calculated parameters at 0\,GPa with the experimental value in~\cite{24}. It is found that the calculated in-plane lattice constant at 0\,GPa is just slightly underestimated by 1.7\% as compared to the experimental value of 4.489 \AA. This indicates that our calculations can reproduce the crystal structures of BaPd$_2$As$_2$ well. Figure 6a shows the total density of states of unit cell BaPd$_2$As$_2$, the region around the Fermi level is shown by the zoomed spper inset. The electronic states at the Fermi level decrease as pressure is increased, this decreasing is microscopically derived from the decrease of partial densities of states of Pd-4$d$ and As-4$p$ as presented in Fig.\,6b. We compare the density of states of BaPd$_2$As$_2$ with those in CaPd$_2$As$_2$~, SrPd$_2$As$_2$ and the non-superconducting CeMg$_2$Si$_2$-type BaPd$_2$As$_2$\cite{25,26}. In all these compounds, the density of states located in the interval from -1\,eV to the Fermi level mainly comes from the contributions of Pd-4$d$ and As-4$p$ orbitals. In addition, the contribution to the valence bands of $A$ ions (Ba, Ca, and Sr) are negligible, indicating the interactions between $A$ ions and -[PdAs]$_2$ blocks are quite ionic. Although the electronic states at the Fermi level in superconducting BaPd$_2$As$_2$ and non-superconducting CeMg$_2$Si$_2$-type BaPd$_2$As$_2$ are both about 2\,states/eV per formula unit, their topology of bands are different, which leads to significant difference of the topology of their Fermi surfaces. Figure 6c shows the 3D Fermi surfaces in the Brillouin zone under pressures of 0 and 5\,GPa. The topology of 3D Fermi surfaces of BaPd$_2$As$_2$~under $P$ = 0 is similar to those in isostructural CaPd$_2$As$_2$~and SrPd$_2$As$_2$ where two electron-like and one quasi-2\,D hole-like Fermi surface sheets are observed~\cite{25,26}. Such a topology is distinguished from that in CeMg$_2$Si$_2$-type BaPd$_2$As$_2$ where a large 3D hole-like sheet and two electron-like sheets are formed in the center and corners of Brillouin zone, respectively. For BaPd$_2$As$_2$ in our study, a quasi-2D hole-like Fermi surface is formed around the Z point at ambient pressure, but this Fermi surface sheet disappears when an external pressure is applied. Therefore, the observed suppression of $T_{c}$ of BaPd$_2$As$_2$ under pressures may also reflect the decrease of density-of-states at the Fermi level.



In order to semi-quantitatively evaluate the change of $T_{c}$ under pressures from first-principles study, we further carried out electron phonon coupling calculations, \textcolor[rgb]{1.00,0.00,0.00}{see Fig.\,7}. The standard approach of evaluating $T_{c}$ is to solve the Eliashberg equation by introducing the energy cutoff and the pseudo Coulomb potential, $\mu^*$\cite{Mc4}.
Here, we adopt the convenient formula of $T_{c}$ that approximates the Eliashberg equation developped by McMillan\cite{Mc5} and Allen-Dynes\cite{Mc6}.
In the McMillan-Allen-Dynes formula, $T_{c}$ is given as
\begin{align}
	T_{c} = \frac{\omega _{log}}{1.2} \exp \left( - \frac{1.04(1+\lambda)}{\lambda - \mu^*(1 + 0.62 \lambda)} \right).
\end{align}
Here, $\mu^*$ is a pseudo Coulomb potential, the electron phonon coupling parameter $\lambda$ and the logarithmic average phonon frequency $\omega _{log}$ are defined as
\begin{align}
	\lambda &= 2 \int d \omega \frac{\alpha^2 F(\omega)}{\omega},\label{eq:lambda}\\
	\ln \omega _{log} &= \frac{2}{\lambda} \int d \omega \frac{\alpha^2 F(\omega)}{\omega} \ln(\omega) \label{eq:omegalog}
\end{align}
with
\begin{align}
	\alpha^2 F(\omega) = \frac{1}{\nef}
	\sum_{n \bm k, m \bm q, \nu} | g_{n \bm k, m \bm k + \bm q}^{\nu} |^2 \delta(\xi_{n \bm k}) \delta(\xi_{m \bm k+ \bm q})
	\delta(\omega - \omega_{\nu \bm q}).
	\label{eq:alpha2F}
\end{align}

Here, $\nef = \sum_{n \bm k}\delta(\xi_{n \bm k})$ is the density of states at the Fermi level, $\xi_{n \bm k}$ is a one-particle band energy with respect to the Fermi level at band index, $n$ and wave vector, $\bm k$, $\omega_{\nu \bm q}$ is the phonon frequency at phonon mode $\nu$ and wave vector, $\bm q$,  and $g_{n \bm k,m \bm k + \bm q}^\nu$ is the electron-phonon coupling.
The simple approach to evaluate Eq. (4) is to take a discrete summation on finite $\bm k$- and $\bm q$-point meshes by replacing two $\delta$ functions with smearing functions with an appropriate smearing width. To get the convergence sufficiently in the first-principles study, a very dense grid of ${32 \times 32 \times 32}$ $\bm k$-mesh and ${4 \times 4 \times 4}$ $\bm q$-point meshes with Methfessel-Paxton smearing width of 0.02 Ry were used in our calculations.

The calculated $T_{c}$ of BaPd$_2$As$_2$ at 0 GPa is 3.1 K, this is in good agreement with the experimentally observed $T_{c}$ of 3.85 K. The corresponding $\lambda$ is 0.70, thus the specfic heat coefficient $\gamma$ can be evaluated  from
\begin{align}
	\gamma = \frac{1}{3} \pi^2 k_B^2 N(E_F)(1+\lambda).\label{eq:shc}
\end{align}
The obtained $\gamma$ is 7.6 mJ/(K$^2\cdot$mol),
which is very close to the experimentally observed $\gamma$ of 6.5 mJ/(K$^2\cdot$mol)\cite{Mc7}.
At prssures of 5 and 10 GPa, $T_{c}$ is decrased to 1.7 K and 1.1 K, respectively; , this is well consistent with the $T_{c} \approx$ 1\,K extrapolated to 10\,GPa based on the measurements up to 9\,GPa  (see inset to Fig.\,5a); $\lambda$ is also reduced to 0.56 and 0.47, respectively. We should notice that the pressure dependence of the density of state has been estimated from the pressure effect on the coefficient of the $T^{2}$ term in the resistivity.

High-pressure synchrotron X-ray diffraction (HPXRD) measurements was performed at room temperature to study the structure robustness and composition integrity of BaPd$_2$As$_2$ upon heavy pressurization. At ambient conditions, BaPd$_2$As$_2$ crystallizes in space group I4/mmm, and its crystal structure is shown in the inset of Fig.\,8d. The pressure-dependent polycrystal X-ray diffraction patterns shown in Fig.\,8a did not reveal any crystallographic symmetry change up to 30.3\,GPa. We refined the ADXRD patterns with GSAS Software based on room temperature polycrystal XRD of BaPd$_2$As$_2$~\cite{xrd}. The refined lattice parameters $a$ = 4.474(2) \AA, $c$ = 10.331(5)\,\AA and $V$ =206.83(4)\,\AA$^3$ are slightly smaller than those at ambient conditions~\cite{24}. The lattice parameters as $a$ function of pressure is plotted are shown in Figs.\,8b and c, where two discontinues can be seen at 12\,GPa and 27.2\,GPa, respectively, which suggests the lattice distortion with pressure. The V-P plot is fitted with a second-order Birch-Murnaghan equation of state (EOS), as shown in Fig.\,8d, which yielded a bulk modulus of $B_{0}$ = 40(6)\,GPa below 12\,GPa and $B_{0}$ = 142(3)\,GPa below 27.2\,GPa. However, the volume collapses above 27.2\,GPa which maybe imply the phase transition. Therefore, these results suggest a subtle structural deformation below 12\,GPa and 27.2\,GPa. The disappearance of superconducting at 11\,GPa may be connected to the large lattice distortion after 12\,GPa, when the compressibility changes dramatically which leads the lattice staying far away from initial structure configuration. As black lines presented in Fig.\,8(a), the upper patterns and lower ones as red lines for the the data taken at increasing and decreasing pressure, respectively. These results show no symmetry change during pressurization.

\section{IV. Conclusion}

In summary, as far as we are aware, no other iron-free layered pnictide superconductor exhibit upper and lower critical fields that behave in the same way as those of BaPd$_2$As$_2$. The difference, we believe, might be linked to this system distinctive Fermi-surface topology. The magnetization hysteresis loops in Fig.\,2 appear to be not far from the borderline to a type-I superconductivity. The present results, (i) the $T$-dependencies of both $H_{c2}$ and $H_{c1}$, (ii) the isotropic single-band alpha model fit to the electronic specific heat, (iii) the large specific-heat jump at $T_{c}$, (iv) the small value of the effective Ginzburg-Landau parameter $\kappa$, (vi) the BCS type temperature dependence of the superconducting energy gap extracted from Andreev reflection resonance, (v) the linear decay of superconductivity under pressure, and (vii) complete pressure suppression of the superconductivity at 12 GPa due to the lattice deformation,  all indicate a strong-coupling conventional isotropic superconductivity in the iron-free layered pnictide superconductor BaPd$_2$As$_2$. Such behaviour differs significantly from all layered superconductors previously known, especially from the unconventional iron-based superconductors with its various competing or coexisting electronic orders.

\begin{acknowledgments}
We are grateful to the support of the DFG, Deutsche Forschungsgemeinschaft through MO 3014/1-1. AS and AU are grateful to the support of the RSCF grant 16-12-10507, VP acknowledges support by the RSCF grant 16-42-01100. HPCAT operations are supported by DOE-NNSA under Award No. DE-NA0001974 and DOE-BES under Award No. DE-FG02-99ER45775, with partial instrumentation funding by NSF. APS is supported by DOE-BES, under Contract No. DE-AC02-06CH11357.
\end{acknowledgments}

\end{document}